\documentclass[twocolumn,showpacs,preprintnumbers,amsmath,amssymb]{revtex4}
\def\Vec#1{\mbox{\boldmath $#1$}}

\usepackage{graphicx}
\usepackage{dcolumn}
\usepackage{bm}

\begin{document}

\draft
\title{Hydrodynamics of driven granular gases}
\author{Hisao Hayakawa}
\address{Department of Physics, Kyoto
University, Kyoto 606-8501}
\begin{abstract}
Hydrodynamic equations for granular gases driven by the Fokker-Planck operator
are derived.
Transport coefficients
 appeared in Navier-Stokes order 
change from the values of a free cooling state to those of a steady
 state. 
\end{abstract}
\pacs{81.05.Rm,47.20.-k 05.20.Dd}
\maketitle

\section{Introduction}

The gas kinetic theory of elastic particles played important roles in 
history of statistical mechanics\cite{boltzmann,chapman}.
When
there are inelastic interactions among particles, the behavior of
collections of particles is completely different from that of elastic
particles: There are no equilibrium states and any spatially homogeneous
states are
no longer stable. 
Such a collection of the particles having inelastic interactions is
called the granular gas whose physical realization can be observed in 
rings of planets, small planets, suspended particles in fluidized beds,
aerosols, and rapid granular flows {\it etc.}\cite{granular-gas}.

A typical example of
granular gases can be found in aerosols  or suspensions
{\it etc.} in which the buoyancy is balanced with gravity.
\cite{friedlander,tanaka}  
In dense suspensions
the hydrodynamic interaction among particles are important\cite{ichiki}, but 
effects of the air may be regarded as a thermostat driven by
the Langevin forces in dilute 
suspensions. Most of researches for granular gases are interested in
 undriven systems which are difficult to be achieved in actual experiences.
We believe that more systematic studies for driven granular gases are required.
Montanero and Santos 
 analyzed statistical properties of the homogeneous state and
 hydrodyanmics of granular gases the white noise
thermostat and the Gaussian thermostat.\cite{montanero,santos02}
Carrilio {\it et al.}\cite{cercignani} and
 Pagnani {\it et al.}\cite{pagnani} analyzed granular 
gases in driven systems by the Langevin force which includes 
both
the white noise with the friction force 
in proportion to the velocity of
particles, but they are not interested 
in hydrodynamics of such the granular fluid.
Since we believe that the model driven 
by such the thermostat, we call the Langevin thermostat,
can describe physical situations of dilute suspensions, 
we need to clarify  properties of the hydrodynamics
of such the system. 
The main purposes of this paper is to investigate
 the effects of Langevin 
thermostat for transport coefficients

The organization of this paper is as follows. In the next section, we
introduce our model and the framework of the Chapman-Enskog method for
the analysis of granular gases. In section III, we calculate the
valuables such as the granular temperature and the fourth cumulant 
in a homogeneous state. In section IV, we obtain the transport
coefficients such as the viscosity and the heat conductivity. In section 
V, we discuss and summarize our results.

\section{Framework}

We consider rarefied gases of smooth identical particles with the
mass $m$, the velocity ${\bf v}$ and the diameter $\sigma$.
The distribution function $f({\bf r},{\bf v}, t)$ 
in our system obeys the
inelastic Boltzmann equation
\begin{equation}\label{b1}
(\partial_t+{\bf v}\cdot\nabla)f=J[f,f]+L_{FP}f,
\end{equation}
where $J[f,f]$ represents the collisional integral 
 given by\cite{goldstein,noije98}
\begin{eqnarray}\label{hardcore}
J[f,h]&=&
\sigma^{d-1}\int d\hat{\Vec \sigma}\int d{\bf v}_1
\Theta({\bf g}\cdot\hat{\Vec\sigma}){\bf g}\cdot\hat{\Vec\sigma}
\nonumber \\
& & \times
(e^{-2}b^{-1}-1)f({\bf r},{\bf v},t)h({\bf r},{\bf v}_1,t)
\end{eqnarray}
where $\Theta(x)$ is Heviside's function,
${\bf g}={\bf v}-{\bf v}_1$, and  
$\Vec{\hat\sigma}$ as the unit vector along the line
connecting centers of mass of contacting particles. The operator
$b^{-1}$ is the inverse of the collisional operator $b$ which are
defined as
\begin{eqnarray}
\label{b3}
b {\bf g}
&=&
{\bf g}-(1+e)({\bf g}\cdot \Vec{\hat\sigma})\Vec{\hat\sigma} ,
\\
b^{-1} {\bf g}
&=&
{\bf g}-
\frac{1+e}{e}({\bf g}\cdot \Vec{\hat\sigma})\Vec{\hat\sigma},
\label{b5}
\end{eqnarray}
where $e$ is the coefficient of restitution which is ranged $0<e\le 1$.
Here we assume that $e$ is a constant for the simplification of our
argument, though the actual coefficient of restitution depends on the
impact velocity\cite{impact,johnson,impact2}.
The effects of the impact velocity dependence of $e$
to macroscopic hydrodynamics can be seen in ref.\cite{brilliantov02}.  
It should be noted that the oblique impacts have important contributions
in actual inelastic collisions.\cite{labous,kuninaka}
We, however, assume that the effects of inelastic oblique collision can be
neglected, which may be justified when particles are smooth hard-core
particles.

The Fokker-Planck operator $L_{FP}$ in eq.(\ref{b1})
represents the driven force coming from the Langevin force as
\begin{equation}\label{FP1}
L_{FP}=\gamma_0\frac{\partial}{\partial {\bf v}}\cdot[{\bf V}
+\frac{T_B}{m}\frac{\partial}{\partial {\bf v}}],
\end{equation}
where the first term and the second term represent the frictional force and 
the thermal activation, respectively. In general, the temperature of the heat 
bath $T_B$ is different from granular temperature $T$. 
If there is no contribution from the collisional integral, the
distribution function is 
relaxed to an equilibrium state as
$f\to f_{eq}\propto \exp[-m V^2/2T_B]$. 
The Fokker-Planck operator (\ref{FP1}) may be categorized into one of the
white noise thermostat
$L_{white}=\gamma_0T_B/m(\partial^2/\partial {\bf v}^2).$
However, 
the contribution of $L_{FP}$ to 
the distribution function, the transport
coefficients and stability of the homogeneous state is different from
the pure white noise thermostat.
We also note that the frictional force in the first term of (\ref{FP1})
has the inverse sign of the Gaussian thermostat\cite{montanero}
$L_{Gauss}=-\gamma_0(\partial/\partial {\bf v}) {\bf V}$
which is also used in simulation of molecular hydrodynamics.\cite{hoover}

Hydrodynamic variables to characterize the macroscopic behavior of  
the gas are the number density, the velocity field and the granular 
temperature 
defined by
\begin{eqnarray}
n({\bf r},t)&=& \int d{\bf v}f({\bf r},{\bf v}, t), \\
n({\bf r},t){\bf u}({\bf r},t)&=& \int d{\bf v} {\bf v}f({\bf r},{\bf v}, t) , \\
\frac{d}{2}n({\bf r},t) T({\bf r},t)&=& 
\int d{\bf v}\frac{1}{2}m {\bf V}^2 f({\bf r},{\bf v}, t),
\label{temp}
\end{eqnarray}
where ${\bf V}\equiv {\bf v}-{\bf u}$.
The integral of the collisional invariance multiplied by $J[f,f]$ over
${\bf V}$ is zero.
Since the loss of kinetic energy in each collision
is given by
\begin{equation}\label{bloss}
\Delta E=-\frac{1-e^2}{4}m ({\bf g}\cdot \Vec{\hat\sigma})^2,
\end{equation}
the following relation holds
\begin{equation}\label{b6}
 \int d{\bf v} 
\frac{1}{2}m V^2 
 J(f,f)
=
-\frac{nd}{2}T \zeta[f,f]
\end{equation}
Here, the cooling rate $\zeta$ in eq.(\ref{b6})can be evaluated 
approximately\cite{noije98}
\begin{equation}\label{zetaH}
\zeta\simeq \zeta^{(0)}=\frac{1-e^2}{4d}\nu_0(d+2)(1+\frac{3}{16}{a_2}),
\end{equation}
where 
$\nu_0=
\pi^{-1/2}n\sigma^{d-1}(T/m)^{1/2}4\Omega_d/(d+2)$,
${a_2}$ is the forth cumulant  defined by
\begin{eqnarray}\label{a2-cum1}
a_2&\equiv& \frac{d}{d+2}\frac{<V^4>}{<V^2>^2}-1, \\
 <V^k>&\equiv& \frac{1}{n}\int d{\bf V}V^k f({\bf V},t)
\end{eqnarray}
which will be determined later.


The balance equations for hydrodynamic variables are
\begin{eqnarray}\label{b8}
D_t n&+& n\nabla\cdot {\bf u}=0, \\
\label{b9}
 D_t u_i&+&(mn)^{-1}\nabla_jP_{ij}=0, \\
\label{b10}
 D_t T&+& \frac{2}{dn}(P_{ij}\nabla_j u_i+\nabla\cdot {\bf q})+T\zeta
\nonumber \\
&= &2\gamma_0
(T_B-T),
\end{eqnarray}
where $D_t=\partial_t+{\bf u}\cdot\nabla$.
The pressure tensor $P_{ij}$ and the heat flux ${\bf q}$ are
respectively defined by
\begin{eqnarray}
P_{ij}&=& m \int d{\bf v} V_iV_j f({\bf r},{\bf v},t) ,
\\
{\bf q}&=& \frac{m}{2}\int d{\bf v} V^2{\bf V} f({\bf r},{\bf v},t).  
\end{eqnarray}

We adopt the Chapman-Enskog method\cite{chapman}, where
space and time dependences appear through
hydrodynamic variables. 
The expansion parameter is regarded as the magnitude of spatial
inhomogeneity. Thus, we expand $f$ around the homogeneous solution
$f^{(0)}$ as 
$f=f^{(0)}+\varepsilon f^{(1)}+\varepsilon^2f^{(2)}+\cdots, $
derivative is also expanded as $\partial_t={\partial_t}^{(0)}+\varepsilon
{\partial_t}^{(1)}+\cdots$. 
Here we apply the Chapman-Enskog method for dilute
granular gases 
developed by 
Brey {\it et al.}\cite{brey98,brey00} and Santos\cite{santos02}
to driven systems. 
To remove the ambiguity of the distribution function, we impose
the solubility conditions in which hydrodynamic variables are 
unchanged from the evaluation by $f^{(0)}({\bf v},t)$.


\section{Homogeneous states}


As the first step of Chapman-Enskog method, we need also 
to obtain the homogeneous
solution of the inelastic Boltzmann equation.
We usually assume the scaling form
\begin{equation}\label{b16_1}
f({\bf v},t)=n v_0(t)^{-d}
\tilde f({\bf c},\tau), \quad {\bf c}={\bf V}/v_0(t)
\end{equation}
with $d\tau=\omega_E dt$ and $v_0(t)=\sqrt{2T/m}$.
The forth cumulant introduced in (\ref{a2-cum1}) for the scaling
function is related to $<c^4>\equiv \int d{\bf c} c^4\tilde f({\bf c})$ as
$<c^4>= d(d+2)(a_2+1)/4$.
Here $\omega_E$ is Enskog's collision frequency given by
\begin{equation}
\omega_E=\frac{d+2}{4}\nu_0=
\displaystyle\sqrt{\frac{2}{\pi}}\Omega_dn\sigma^{d-1}v_0.
\end{equation}

For the calculation
we need to obtain 
\begin{equation}
\mu_k\equiv -\int d{\bf c} c^k J[\tilde f,\tilde f]
\end{equation} 
The cumulants and $\mu_k$ 
can be evaluated by an
approximate expansion of Sonine polynomials\cite{noije98}. 
For example, $\mu_2$
is evaulated as
\begin{equation}
{\mu_2}\simeq \frac{1}{2}(1-e^2)\frac{\Omega_d}{\sqrt{2\pi}}
(1+\frac{3}{16}{a_2}),
\end{equation}
$\mu_4$  is estimated as \cite{noije98} 
\begin{equation}  
{\mu_4}\simeq \displaystyle\sqrt{\frac{2}{\pi}}\Omega_d
(A_1+{a_2}^H A_2),
\end{equation}
with
\begin{eqnarray}
A_1&=& \frac{1-e^2}{4}(d+\frac{3}{2}+e^2) \nonumber \\
A_2&=& \frac{3}{128}(1-e^2)(10d+39+10e^2) \nonumber \\
& & +\frac{1}{4}(1+e)(d-1).
\end{eqnarray}
It should be noted that these evaluations are based on two
approximations: (i) the truncation of the first Sonine expansion, and
(ii) the linearization of $a_2$. If we adopt the first assumption, 
 the linearization of $a_2$ gives a nice 
evaluation,\cite{brilliantov} (iii) the direct comparison of the
transport coefficients obtained by the Monte
Carlo simulation and the linearized approximation gives good agreement
in free cooling states.\cite{brey00} 
 However, nobody knows how to converge
the Sonine expansion for driven granular gases. Thus, we may need to check the
convergence of the Sonine expansion as in the case of elastic
particles.\cite{kim}

Now, let us discuss the time evolution of temperature field.
Equation (\ref{b10}) becomes
\begin{equation}\label{b10a1}
{\partial_t}^{(0)}\theta=2\gamma_0-(2\gamma_0+\zeta)\theta
\end{equation}
in the homogeneous state with the Langevin thermostat, where
$\theta\equiv T/T_B$.
Assuming $\theta=\theta^{(0)}+a_2\theta^{(1)}+O({a_2}^2)$,
eq.(\ref{b10a1}) is reduced to
\begin{equation}\label{b10a2}
{\partial_{\tau}}\theta^{(0)}=2\{\hat\gamma-(\hat\gamma+\hat\zeta)\theta^{(0)}\} 
\end{equation}
in the lowest order, where 
$\hat\gamma=\gamma_0/\omega_E$ and 
$\hat\zeta=(1-e^2)/2d$.
This equation has the solution
\begin{equation}\label{b10a3}
\theta^{(0)}=\theta_{\infty}+
(\theta^{(0)}(0)-\theta_{\infty})e^{-2(\hat\gamma+\hat\zeta)\tau}
\end{equation}
with $\theta_{\infty}\equiv \hat\gamma/(\hat\gamma+\hat\zeta)$.

In the scaling limit, the inelastic Boltzmann equation (\ref{b1}) is
reduced to
\begin{eqnarray}\label{inel-Bol}
\frac{\Omega_d}{\sqrt{2\pi}}\partial_{\tau}\tilde f({\bf c},\tau)
&=&
\tilde J[\tilde f,\tilde f]+
\left(\frac{\hat\gamma}{\theta}-\frac{\mu_2}{d}\right)
\frac{\partial}{\partial {\bf c}}\cdot({\bf c}\tilde f({\bf c},\tau))
\nonumber \\
& &+\frac{\hat\gamma}{2\theta}\frac{\partial^2}{\partial c^2}\tilde f({\bf c},\tau)
\end{eqnarray}
where $J[f,f]=n^2{(\sigma/v_0(t))}^{d-1}\tilde J[\tilde f,\tilde f]$.
From the equation for $<c^4>=d(d+2)(a_2+1)/4$
we obtain the equation of lowest order of $a_2$:  
\begin{eqnarray}\label{b10a3.5}
\partial_{\tau}a_2&=&4\hat\zeta-\hat A_1+a_2[\frac{19}{4}\hat\zeta
-4\frac{\hat\gamma}{\theta}-\hat A_2]+\frac{3}{4}\hat\zeta {a_2}^2 \\
&\simeq &4\hat\zeta-\hat A_1+a_2[\frac{19}{4}\hat\zeta
-4\frac{\hat\gamma}{\theta^{(0)}}-\hat A_2]
\label{b10a4}
\end{eqnarray}
with $\hat A_1=8A_1/d(d+2)$ and $\hat A_2=8A_2/d(d+2)$.
The solution of (\ref{b10a4}) is given by
\begin{eqnarray}\label{b10a5}
a_2(\tau)&\simeq& {a_2}^{\infty}+(a_2(0)-{a_2}^{\infty}) \nonumber \\
& & \times
\exp[(\frac{19}{4}\hat\zeta-\hat A_2)\tau-4\hat\gamma\int_0^{\tau}\frac{d\tau'}{\theta^{(0)}(\tau')}]
\end{eqnarray}
where
\begin{equation}
{a_2}^{\infty}=\frac{\hat A_1-4\hat\zeta}{\frac{19}{4}\hat\zeta-4(\hat\gamma+\hat\zeta)-\hat A_2} .
\end{equation}
Although we present the result of linearization of $a_2$, it is possible 
to obtain the exact steady values of $\theta$ and $a_2$ for
(\ref{b10a1}) and (\ref{b10a3.5}). The result is presented in Appendix,
but the differences between the exact values and the result from
linearized approximation are invisible (Fig.1). 
We also  compare  $a_2$ in eq.(\ref{b10a5})and $\theta$ in
(\ref{b10a3}) with the numerical solution of (\ref{b10a3.5}) and
(\ref{b10a1}) to confirm the validity of the linearization of $a_2$.

\begin{figure}
 \begin{center}
  \includegraphics[width=78mm]{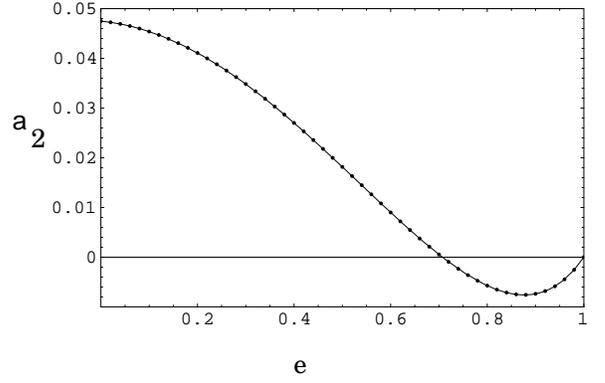}
 \end{center}
\caption{
Steady $a_2$ as a function of $e$ for $d=3$ and $\hat\gamma=0.1$.
The solid line and the solid circles represent the result of linearized
 approximation and the exact one, respectively.
}
\end{figure}

\section{The determination of the transport coefficients}

Here, we explicitly
obtain the result of the transport coefficients for heated granular
gases.
The solution $f^{(0)}$ is isotropic so that the zeroth order pressure and
the heat flux are given by
\begin{equation}\label{b21}
{P_{ij}}^{(0)}=p\delta_{ij}, \quad {\bf q}^{(0)}={\bf 0}
\end{equation}
where $p=n T$ is the hydrostatic pressure.

The first order equation of the Boltzmann equation becomes
\begin{eqnarray}\label{b22}
({\partial_t}^{(0)}+L-L_{FP})f^{(1)}&=&-({\partial_t}^{(1)}+{\bf v}\cdot\nabla)f^{(0)} \nonumber \\
&=&-({D_t}^{(1)}+{\bf V}\cdot\nabla)f^{(0)}
\end{eqnarray}
with ${D_t}^{(1)}={\partial_t}^{(1)}+{\bf u}\cdot\nabla$. Here 
the linear operator $L$ in (\ref{b22}) is defined by
\begin{equation}\label{b23}
Lf^{(1)}=-J[f^{(0)},f^{(1)}]-J[f^{(1)},f^{(0)}] .
\end{equation}


Multiplying both side of (\ref{b22}) by $m V_i V_j$ and integrate
over ${\bf V}$, we obtain
\begin{equation}\label{s47}
({\partial_t}^{(0)}+\nu){P_{ij}}^{(1)}+{\Pi_{ij}}^{(1)}=-p 
\Delta_{ijkl}\nabla_k u_l
\end{equation}
where
\begin{eqnarray}\label{s48}
{\Pi_{ij}}^{(1)}&\equiv& -m\int d{\bf V}V_i V_j L_{FP}f^{(1)}, \\
{\Delta_{ijkl}}&\equiv& \delta_{ik}\delta_{jl}+\delta_{il}\delta_{jk}
-\frac{2}{d}\delta_{ij}\delta_{kl}.
\end{eqnarray}
The solution of (\ref{s47}) can be written as
\begin{equation}\label{b28}
{P_{ij}}^{(1)}=-\eta\Delta_{ijkl}\nabla_k u_l
\end{equation}
where $\eta$ is the viscosity. 

It is possible to obtain $\nu$ in eq.(\ref{s47}) through the relation
\begin{equation}\label{s40}
m \int d{\bf V} V_iV_j Lf^{(1)}({\bf V})=\nu {P_{ij}}^{(1)}.
\end{equation}
The evaluation of $\nu$ is independent of the existence of the
thermostat. We have evaluated  $\nu$ as\cite{noije98}:
\begin{equation}\label{sB6H}
{\nu_{\eta}}^*\equiv\frac{\nu}{\nu_0}\simeq \frac{3}{4d}(1-e+\frac{2}{3}d)(1+e)
(1-\frac{1}{32}{a_2}).
\end{equation}

With the aid of
\begin{equation}
{\partial_t}^{(0)}{P_{ij}}^{(1)}=\left(\frac{\gamma_0}{\theta}
-\gamma_0-\frac{\zeta}{2}\right){P_{ij}}^{(1)},
\end{equation}
eqs.(\ref{lt1}),(\ref{lt2}) and (\ref{s47}) lead to
\begin{equation}\label{lt3}
\eta^*\equiv \frac{\eta}{\eta_0}=\frac{2\gamma^*+1}{
\gamma^*(1+\theta^{-1})+{\nu_{\eta}}^*-\zeta^*/2}
\end{equation}
where $\gamma^*=\gamma/\nu_0$, ${\nu_{\eta}}^*=\nu/\nu_0$ and
$\zeta^*=\zeta/\nu_0$. $\eta_0(\hat\gamma)$ 
is the viscosity for $e=1$ which is different from the value $\eta_e$ of the
elastic gas as $\eta_{el}=\eta_0(2\hat\gamma^*+1)$. 
The steady value of $\eta^*$ is obtained when we substitute
$\theta^{\infty}$ and ${a_2}^{\infty}$ into (\ref{lt3}).

Let us consider the heat flux. Multiplying both side of
 eq.(\ref{b22}) by
$m V^2{\bf V}/2$ 
and integrate over ${\bf V}$ we obtain
\begin{eqnarray}\label{s54}
({\partial_t}^{(0)}+\nu'){\bf q}^{(1)}+{\bf Q}^{(1)}&=&-\frac{d+2}{2}(1+2a_2)\frac{p}{m}\nabla T \nonumber \\
& & -\frac{d+2}{2}a_2\frac{T^2}{m}\nabla n,
\end{eqnarray}
where
\begin{equation}\label{q1}
{\bf Q}^{(1)}=-\frac{m}{2}\int d{\bf V}V^2{\bf V}L_{FP}f^{(1)}.
\end{equation}
Here, $\nu'$ has already been calculated
as\cite{brey00} 
\begin{eqnarray}\label{nuH'}
{\nu_{\kappa}^*}\equiv\frac{{\nu}'}{\nu_0}&\simeq& \frac{1+e}{d}[
\frac{d-1}{2}+\frac{3}{16}(d+8)(1-e) \nonumber \\
& &+\frac{4+5d-3(4-d)e}{512}{a_2}.
]
\end{eqnarray}

The heat flux is described by
\begin{equation}\label{s58}
{\bf q}^{(1)}= -\kappa \nabla T-\mu \nabla n,
\end{equation}
where $\eta$ and $\kappa$ are the shear viscosity and the thermal 
conductivity, respectively. The other transport coefficient $\mu$ appears
only is granular gases.
Through the substitution of (\ref{s58}) and (\ref{nuH'}) 
with the result of ${\bf Q}^{(1)}$ 
into (\ref{s54}) we can obtain $\kappa$ and $\mu$.

From the scaling form (\ref{b16_1})
the following relations should be satisfied:
\begin{eqnarray}\label{lt1}
{\partial_t}^{(0)}{\bf q}^{(1)}&=&
(2\zeta+3\gamma_0-\gamma_0/\theta)\kappa\nabla T \nonumber \\
& &+\{\mu[\frac{3}{2}\zeta+3\gamma_0(1-\theta^{-1})]+\frac{\kappa \zeta T}{n}\}
\nabla n
.
\end{eqnarray}
On the other hand,  ${\Pi}^{(1)}$ and ${\bf
Q}^{(1)}$ have the relations
\begin{equation}\label{lt2}
{\Pi_{ij}}^{(1)}=2\gamma_0 {P_{ij}}^{(1)}, \quad
{\bf Q}^{(1)}=3\gamma_0 {\bf q}^{(1)}.
\end{equation}
Thus, $\kappa$ and $\mu$ can be obtained from
(\ref{lt1}),(\ref{lt2}) and (\ref{s54}) as
\begin{eqnarray}\label{lt6}
\kappa&=&\frac{d+2}{2}\frac{1+2a_2}{\nu'-2\zeta+\gamma_0/\theta}
\frac{ n T}{m}, \\
 \mu &=& \frac{\frac{\kappa \zeta T}{n}+\frac{d+2}{2}a_2\frac{T^2}{m}}
{\nu'-\frac{3}{2}\zeta-3\gamma_0/\theta}.
\end{eqnarray}
Thus, $\kappa^*=\kappa/\kappa_0$ which is the normalized heat
conductivity by its value of $e=1$, $\kappa_0(\hat\gamma)$, is given by
\begin{eqnarray}\label{lt7}
{\kappa^*}
&=& (\frac{d-1}{d}+\gamma^*)\frac{1+2a_2}{{\nu_{\lambda}}^*-2\zeta^*+\gamma^*/\theta} \\
\mu^*&= & 
\frac{\kappa^*}{1+2a_2}
\frac{a_2({\nu_{\kappa}}^*+\gamma^*/\theta)}
{{\nu_{\kappa}}^*-\frac{3}{2}\zeta^*+3\gamma^*/\theta}
\end{eqnarray}
Their steady values are evaluated replacing $\theta$ and $a_2$ by their steady 
values.

Figure 2 shows the time evolution of $\eta^*$, $\kappa^*$ and $\mu^*$
for $e=0.9$, $d=3$, $\hat\gamma=0.1$. Figure 3 shows the steady values
of $\eta^*$, $\kappa^*$ and $\mu^*$ as functions of $e$ for $d=3$ and
$\hat\gamma=0.1$. As we can see from figures, $\mu^*$ can be comparable
with others. This situation has not been realized for free cooling systems.

It is also remarkable that  $\mu^*$ is not proportional to $a_2$ and
keeps positive for the region of $a_2<0$ even in the
steady limit. This result is contradicted with the framework by 
Santos\cite{santos02}.
The difference comes from the following. In our system, the system is
heated by $T_B$ uniformly, but the temperature field $T$ contains
spatial fluctuations. Thus, ${\partial_t}^{(0)}{\bf q}^{(1)}$ cannot be
zero even in steady states, because this contains the term in proportion to
$\nabla[2\gamma_0T_B+(2\gamma_0+\zeta)T]$.

\begin{figure}
 \begin{center}
  \includegraphics[width=78mm]{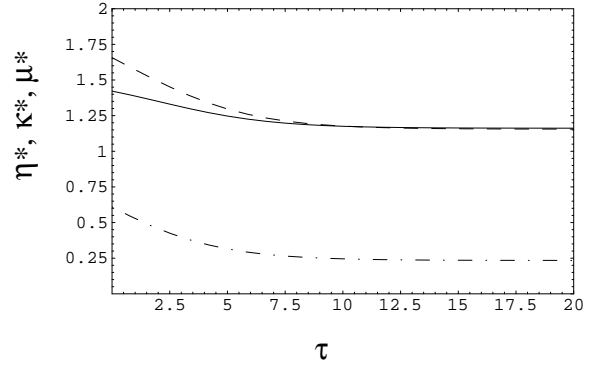}
 \end{center}
\caption{
The time evolution of $\eta^*$ (solid line), $\kappa^*$ (dashed line)
 and $\mu^*$ (dot-dashed line) for $e=0.9$, $d=3$, $\hat\gamma=0.1$
and $\theta(0)=2$.
}
\end{figure}

\begin{figure}
 \begin{center}
  \includegraphics[width=78mm]{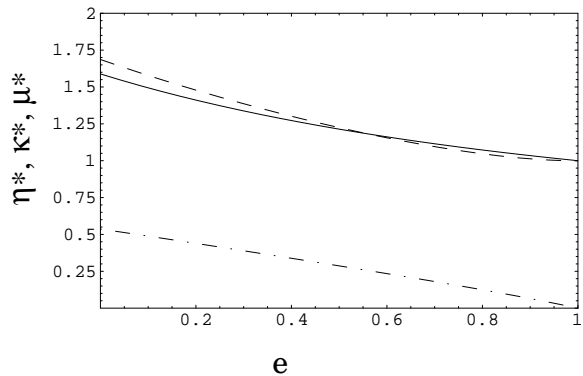}
 \end{center}
\caption{
The steady values of $\eta^*$ (solid line), $\kappa^*$ (dashed line)
 and $\mu^*$ (dot-dashed line) for $e=0.9$, $d=3$, $\hat\gamma=0.1$.
}
\end{figure}

\section{Concluding Remarks}

We have derived hydrodynamic equations 
based on a systematic Chapman-Enskog method for dilute granular gases 
driven by the Langevin thermostat in this paper.  
We have determined 
all of the transport coefficients $\eta$, $\kappa$ and $\mu$ 
 appear in Navier-Stokes order as a function of the 
 restitution coefficient $e$. 

The result is based on the linearized approximation of $a_2$ in the first order
truncation of Sonine expansion. Although we believe that this
approximation gives a nice evaluation, nobody knows its theoretical
background and quantitative validity. In particular, Pagnani {\it et
al.}\cite{pagnani} has reported that deviation of $f$ from the Gaussian
in driven granular gases  is large. So we may need to check the
convergence of the Sonine expansion and to compare the theoretical
prediction with simulations. We also need to look for the possibility
to apply our result to explain the data in actual experiments for
suspensions.

\vspace*{0.5cm}

The author would like to thank A. Santos for fruitful discussion. 
This study is partially supported by the Inamori Foundation.

\appendix
\section{Steady values of $a_2$ and $\theta$ }

As mentioned in the text, it is possible to obtain exact steady
solutions of (\ref{b10a1}) and (\ref{b10a3.5}). The result is so 
complicated and difference between them and the linearized solutions is
small that we do not  use the exact form for later discussion.
Here we present the exact solutions:
\begin{equation}\label{appen1}
{a_2}^{\infty}=\frac{16(1-3e^2+2e^4)}{Q_1}
\end{equation}
where $Q_1=73-32e-75e^2-30e^4+64d^2\hat\gamma+8d(7+4e-3e^2+16\hat\gamma)$,
and
\begin{equation}\label{appen2}
\theta_{\infty}=\frac{d\hat \gamma Q_1}{Q_2+64d^3\hat \gamma^2+Q_3-d(1+e)Q_4},
\end{equation}
where $Q_2=2(1+e)^2(19-46e+33e^2-12e^3+6e^4)$, $Q_3=8d^2\hat\gamma(11+4e-7e^2+16\hat \gamma)$ and 
$Q_3=-28+e^2(28-30\hat\gamma)-137\hat\gamma+6e^3(5\hat\gamma-2)+e(12+169\hat\gamma)$.


\begin{references}
\bibitem{boltzmann} L. Boltzmann, Lecture Notes on Gas Theory (Dover,
 New York, 1995).
\bibitem{chapman} S. Chapman and T. G. Cowling, The Mathematical Theory
of Nonuniform Gases ; Third Edition (Cambridge Univ. Press, Cambridge, 1970).

\bibitem{granular-gas} T. P\"oschel and S. Luding eds. 
Granular Gases (Springer,  Berlin, 2000).

\bibitem{friedlander} S. K. Friedlander, Smoke, Dust and Haze:
 Fundamentals of Aerosol Behavior (John Wiley $\&$ Sons, New Yprk, 1977).
\bibitem{tanaka} 
T. Kawaguchi, T. Tanaka and Y.Tsuji, Powder Technol. 
{\bf 96}, 129 (1998); T. Kawaguchi, M. Sakamoto, T. Tanaka and Y. Tsuji,
 Powder Technol. {\bf 109}, 3 (2000).

\bibitem{ichiki} K. Ichiki and H. Hayakawa, Phys. Rev. E {\bf 52}, 658 (1995).

\bibitem{montanero} J. M. Montanero and A. Santos, Granular Matter, {\bf
 1}, 57 (1998).
\bibitem{santos02} A. Santos, Physica A {\bf 321}, 442 (2003).


\bibitem{cercignani} J. A. Carrillo, C. Cercignani and I. M. Gamba,
Phys. Rev. E {\bf 62}, 7700 (2000).


\bibitem{pagnani} R. Pagnani, U. M. Bettolo Marconi and A. Puglish,
 Phys. Rev. E {\bf 66}, 051304 (2002).



\bibitem{goldstein}A. Goldstein  and M. Shapiro,
     J. Fluid
     Mech. {\bf 282}, 75 (1995).
\bibitem{noije98} T. P. J. van Noije and M. H. Ernst, Granular Matter
 {\bf 1}, 57 (1998).
\bibitem{impact} W. J. Stronge, Impact Mechanics (Cambridge Univ. Press, 
 Cambridge, 2000).
\bibitem{johnson} K. L. Johnson, Contact Mechanics (Cambridge
 Univ. Press, Cambridge, 1985).
\bibitem{impact2} G. Kuwabara and K. Kono, {\it Jpn. J. Appl. Phys.},
 {\bf 26},  1230 (1987); N. V.  Brilliantov, F. Spahn, J. -M.
 Hertzsch, and T.  P\"oschel,  Phys. Rev.  E {\bf 53},  5382 (1996);
W. A. M.  Morgado and I. Oppenheim, Phys. Rev.  E {\bf 55}, 1940 (1997);
F. Gerl and  A. Zippelius, Phys. Rev. E {\bf 59} , 2361 (1999);
H. Hayakawa and H. Kuninaka, Chem. Eng. Sci. {\bf 57}, 239 (2002).
\bibitem{brilliantov02} N. V. Brilliantov and T. P\"oschel, 
Phil. Trans. R. Soc. Lond. A {\bf 360}, 415 (2002).

\bibitem{labous} L. Labous, A.D. Rosato and R. N. Dave, Phys. Rev. E {\bf 56}, 5717 (1997).
\bibitem{kuninaka} H. Kuninaka and H. Hayakawa, cond-mat/0301483 (to be
 published in J. Phys. Soc. Jpn.).

\bibitem{hoover} W. G. Hoover, Computational Statistical Mechanics
 (Elservier, Amsterdam, 1991).

\bibitem{brey98} J. J. Brey, J. W . Dufty, C. S. Kim and A. Santos,
Phys. Rev. E {\bf 58}, 4638 (1998).
\bibitem{brey00} J. J. Brey and D. Cubero, in ref.\cite{granular-gas}.


\bibitem{brilliantov} N. V. Brilliantov and T. P\"oschel, 
Phys. Rev. E {\bf 61}, 2809 (2000).

\bibitem{kim} H.-D. Kim and H. Hayakawa, cond-mat/0202003.


\end{references}
\end{document}